\documentstyle[12pt,psfig]{article}
\begin{document}
\setlength{\headheight}{0in}
\setlength{\headsep}{0in}
\setlength{\topskip}{1ex}
\setlength{\textheight}{8.5in}
\setlength{\topmargin}{0.5cm}
\setlength{\baselineskip}{0.24in}
\catcode`@=11   
\long\def\@caption#1[#2]#3{\par\addcontentsline{\csname
  ext@#1\endcsname}{#1}{\protect\numberline{\csname
  the#1\endcsname}{\ignorespaces #2}}\begingroup
    \small
    \@parboxrestore
    \@makecaption{\csname fnum@#1\endcsname}{\ignorespaces #3}\par
  \endgroup}
\catcode`@=12
\def\slashchar#1{\setbox0=\hbox{$#1$}           
   \dimen0=\wd0                                 
   \setbox1=\hbox{/} \dimen1=\wd1               
   \ifdim\dimen0>\dimen1                        
      \rlap{\hbox to \dimen0{\hfil/\hfil}}      
      #1                                        
   \else                                        
      \rlap{\hbox to \dimen1{\hfil$#1$\hfil}}   
      /                                         
   \fi}                                         %
\newcommand{\newc}{\newcommand}
\def\be{\begin{equation}}
\def\ee{\end{equation}}
\def\bea{\begin{eqnarray}}
\def\eea{\end{eqnarray}}
\def\simlt{\stackrel{<}{{}_\sim}}
\def\simgt{\stackrel{>}{{}_\sim}}
\begin{titlepage}

\vskip 2cm
\begin{center}
{\Large\bf  
$gg\rightarrow\gamma \bar{f}f$ in the strongly interacting phase of the MSSM}
\vskip 1cm
{\large
 D. A. Demir\footnote{Present Address: High Energy Section, ICTP, Trieste, Italy\\}}
\vskip 0.5cm
{\setlength{\baselineskip}{0.18in}
{\normalsize\it Middle East Technical University, Department of Physics,
06531, Ankara, Turkey\\} } 
\end{center}
\vskip .5cm
\begin{abstract}
$g g \rightarrow \gamma \bar{f}f$ scattering is discussed in the strongly interacting phase of
the MSSM. The rate for the decay $h \rightarrow \gamma \bar{f} f$ is computed in the MSSM and SM,
and values of the Higgs--sfermion coupling needed for the former to dominate on the latter
are identified. It is found that the MSSM signal dominates on the SM one for Higgs--sfermion
couplings well below the one needed for developing stopponium bound states via Higgs mediation.
\end{abstract}
\end{titlepage}
\newpage
The Minimal Supersymmetric Model contains a total of five Higgs scalars; 
two charged, two CP-even and a CP-odd one. It is known that for most 
of the parameter space allowed by the present experimental data 
\cite{exp}, the MSSM is in the decoupling regime \cite{decoupl} in which 
one of the CP-even scalars, CP-odd scalar and charged scalar are rather  
heavy and almost degenerate in mass, while the mass of the lightest CP-even 
scalar, $h$, assumes its upper bound, $m_{h}\simlt 95- 130$ GeV 
\cite{upper}, depending on the region of the parameter space. In the
decoupling regime the lightest Higgs, $h$, has almost the same 
properties as the SM Higgs boson, and presently its mass is bounded from 
below by $m_{h}\simgt 90$ GeV \cite{exp} by the negative Higgs search at 
LEP2. With this mass bounds, $h$ will be the only Higgs scalar accessible at 
the LHC \cite{lhc}. 

After the ending of LEP2 era, search for the Higgs particle will continue at 
the LHC \cite{lhc}. At the LHC energies the Higgs particle is expected to be 
produced via gluon-gluon fusion $pp\rightarrow gg \rightarrow h$ \cite{fusion}. 
The produced Higgs particle subsequently decays either to $\gamma\gamma$ for 
$M_{Z}\simlt m_{h} \simlt 130\;GeV$ or ($Z Z \rightarrow 2\ell^{+}2\ell^{-}$) 
for $130 GeV\simlt m_{h} \simlt 155 GeV$ \cite{gamma,at-cms}, or 
($W^{+}W^{-}\rightarrow \ell^{+}\nu {\ell^{-}}^{\prime}\bar{\nu}$)
for $155 GeV \simlt m_{h} \simlt 2 M_{Z}$ \cite{dreiner}. With the
huge irreducable $\gamma\gamma$ background, isolation of the narrow $\gamma\gamma$
resonance \cite{at-cms,others} requires the design of high- resolution
detectors \cite{at-cms}. Besides these rare decays of the Higgs 
particle, the diHiggs production process $gg\rightarrow 
\phi_{1}\phi_{2}$ is also crucial for constructing the Higgs sector of the 
model under concern as the trilinear Higgs couplings can be probed 
directly with such two scalar final states \cite{diHiggs}. 

Single isolated photon production at LHC dominated by the 
tree-level transition $g q \rightarrow \gamma q$, will be an important 
test of the QCD at large $p^{\gamma}_{t}$ \cite{prompt}. Moreover, 
$h\rightarrow \bar{f}f$ can be a possible signature of 
the intermediate mass Higgs boson if there is a high $p_{t}$ jet 
accompanying the produced Higgs \cite{tau}. Thus, in addition to the 
two-body decays of the produced Higgs, it seems necessary to 
have a discussion of the three-body decay process  
$g g \rightarrow \gamma \bar{f}f$ whose signature consists of a 
single isolated $\gamma$ and a pair of light  
fermions to which the produced $h$ decays. 
    
Altough the processes mentioned above are highly important for 
discovering the Higgs particle at the LHC, there remains still the 
question of which model this would-be discovered scalar particle 
belongs to. To find at least an indirect evidence for the underlying 
model, one has to exploit those properties of the model not shared by the 
other candidate ones. Following the detailed discussion in 
\cite{decoupl}, one immediately observes that, in the decoupling regime, 
the SM and MSSM differs from each other mainly by the  
existence of the supersymmetric partners of the known SM particles 
in the MSSM particle spectrum. Therefore, in the decoupling regime 
one can obtain, albeit indirect, some manifestations of the 
low-energy supersymmetry. In this sense, that region of the MSSM parameter
space in which the supersymmetric partners of the known fermions couple 
to the Higgs particle strongly may provide a room for obtaining some 
signal of the supersymmetry. Indeed, as recently proposed, when the
stop trilinear coupling becomes large, one faces with new phenomena 
ranging from the sfermion bound states to charge and/or color 
breaking minima \cite{sasha}. This very portion of the entire MSSM 
parameter space can cause certain collision processes to have 
amplified rates which, if observed, can be  taken as an indication 
for the MSSM to be the underlying model. In  fact, recently LHC- 
approved processes $h\rightarrow \gamma\gamma$ and $h\rightarrow g g$ 
have been analyzed in this kind of parameter space \cite{djou}.

In this letter we discuss the process 
$g g \rightarrow h^{*} \rightarrow \gamma (Z^{*}, \gamma^{*} \rightarrow) \bar{f} f$
in that region of the MSSM parameter space in which 
\begin{itemize}
\item the heavy Higgs scalars are much heavier than the $Z$ boson, and 
$|\alpha|\approx |\beta-\pi/2|$ up to corrections ${\cal{O}}(M_{Z}^{2}/M_{A}^{2})$ 
\cite{decoupl}, and 
\item the lightest Higgs $h$ couples to the light stop with a stength as large as the one
needed for developing light stop bound states via Higgs mediation \cite{sasha}.
\end{itemize}

As in $g g \rightarrow \gamma \gamma$ there is a non-negligable background 
represented mainly by the box diagrams. However, if the final state fermions 
are tagged properly together with the detection of the photonic jet it may be easier to
observe this event if it has suffienctly large branching fraction \cite{prompt,tau}. 
Therefore, below we compare the predictions of the MSSM and the SM in analyzing the 
process under concern.

In computing one-loop $h \gamma V^{*}$ ($V=\gamma, Z$) vertex $h \tilde{t}_{1}\tilde{t}_{1}$
($\tilde{t}_{1}$ being the light mass-eigenstate stop) coupling will be denoted 
by $g_{h\tilde{t}}$. When expressed in terms of the basic parameters of the MSSM 
Lagrangian $g_{h\tilde{t}}$ is seen to contain two parts: The D--term contributions 
(proportional to the SU(2) coupling $g$), and F--term and soft breaking contribution 
(proportional top Yukawa coupling and related to the stop left-right mixing mass parameters).
In this sense, large $g_{h\tilde{t}}$ implies automatically large stop left-right mixing 
so that the stop mixing angle becomes maximal $\theta_{\tilde{t}}\approx \pi/4$. The 
expressions for the stop masses, mixings, and $g_{h\tilde{t}}$ can be found, for example, 
in \cite{gamma,hunter}. For convenience we will follow the notation of Weiler and Yuan in 
\cite{gamma}. Moreover we introduce the 'fine structure' constant 
\begin{eqnarray}
\alpha_{\tilde{t}}=\frac{1}{16 \pi} \frac{g_{h\tilde{t}}^{2}}{m_{\tilde{t}_{1}}^{2}}
\end{eqnarray}
where $m_{\tilde{t}_{1}}$ stands for the mass of the light stop. This particular form 
for $\alpha_{\tilde{t}}$ is chosen to suggest the formation of stopponium states via 
light Higgs mediation. As the explicit computations in  \cite{sasha,canad} show
such bound states occur when $\alpha_{\tilde{t}}\simgt 1.7 (m_{h}/m_{\tilde{t}_{1}})$. 
Here the main concern is not on the analysis of such bound states but the critical
value of $\alpha_{\tilde{t}}$ for which the MSSM prediction for the process under 
concern exceeds that of the SM by a given amount. 

The basic Higgs search strategy at the LHC is the observation of gluon--gluon fusion to 
Higgs whose resonance shape, width and subsequent dominant decay mode are of central 
importance. As mentioned at the beginning, the $\gamma\gamma$ decay mode is hard to 
detect, and thus, one generally searches for other decay signatures whose observation 
could be easier. In this context one recalls the recent works \cite{list} which deal with 
the associated production of squarks with $h$. Discussion of the process
 $g g \rightarrow \gamma \bar{f} f$ comprises gluon-gluon fusion to Higgs (requiring 
$\Gamma(h\rightarrow g g)$ followed by the Higgs decay to $\gamma \bar{f}f$ final state.
Unless Higgs comes to its mass-shell the process looses its importance for the LHC 
Higgs search. The expression for $\Gamma(h\rightarrow g g)$ can be found in 
\cite{gamma,hunter,djou}. On the other hand the rate for $h\rightarrow \gamma \bar{f} f$
reads as 
\begin{eqnarray}
R\equiv \frac{\Gamma(\mbox{MSSM}| h\rightarrow \gamma \bar{f} f)}{\Gamma(\mbox{SM}| h\rightarrow \gamma \bar{f} f)}
=\frac{\int_{4 m_{f}^{2}}^{m_{h}^{2}} d s\; A_{\mbox{MSSM}}(s)}{\int_{4 m_{f}^{2}}^{m_{h_{0}}^{2}} d s\; A_{\mbox{SM}}(s)} 
\end{eqnarray}  
in units of the SM rate with $m_{f}$ being the mass of the produced fermion and $\sqrt{s}$ is the invariant mass flow to
$\bar{f}f$ channel. Here $m_{h}$ ($m_{h_{0}}$) is the mass of the lighest Higgs in the MSSM (Higgs mass in the SM), and they
are taken equal in writing $R$. The MSSM integrand $A_{\mbox{MSSM}}(s)$ is given by
\begin{eqnarray} 
A_{\mbox{MSSM}}(s)&=&(1-\frac{s}{m_{h}^{2}})^{3} (1- \frac{4 m_{f}^{2}}{s})^{1/2} \{ \frac{2}{3} (s -m_{f}^{2})[
\frac{|A_{\gamma}(s)|^{2}}{s^{2}}\\&+&2 v_{f} \frac{Re[A_{\gamma}(s) A_{Z}(s)]}{s
(s-M_{Z}^{2})}+(a_{f}^{2}+v_{f}^{2})\frac{|A_{Z}(s)|^{2}}{(s-M_{Z}^{2})^{2}}]\nonumber\\&-&a_{f}^{2} m_{f}^{2}
\frac{|A_{Z}(s)|^{2}}{(s-M_{Z}^{2})^{2}}\}\nonumber
\end{eqnarray}
where $v_{f}=(I_{3}^{f}-2 Q_{f} s_{W}^{2})/(s_{W} c_{W})$ and  $a_{f}=-I_{3}^{f}/(s_{W} c_{W})$ are the vector 
and axial--vector couplings of the $Z$ boson, and $A_{\gamma}$ and $A_{Z}$ are the loop functions decsribing 
$h \gamma \gamma^{*}$ and $h \gamma Z^{*}$ effective vertices, respectively. These vertex formfactors get contributions 
from all the charged particle of the model under concern (SM or MSSM). In the MSSM, one has
\begin{eqnarray}
A_{\gamma ,Z}=A^{W^{\pm}}_{\gamma 
,Z}+A^{f^{\pm}}_{\gamma ,Z}+A^{\chi^{\pm}}_{\gamma ,Z}+A^{H^{\pm}}_{\gamma ,Z}+A^{\tilde{f}^{\pm}}_{\gamma ,Z}
\end{eqnarray}
representing, respectively, the loops of $W$ --boson, charged fermions, charginos, charged Higgs boson, and charged sfermions.
$A_{\mbox{SM}}$ in (2) can be obtained from (3) by keeping only $W$--boson and charged fermion contributions in (4). The 
explicit expressions for the loop functions $A^{i^{\pm}}_{\gamma,Z}$ can be found in \cite{gamma}. As described there,
there are three independent loop functions determining $A^{i^{\pm}}_{\gamma,Z}$: $I[m_{loop}^{2},
m_{h}^{2}, s]$, $J[m_{loop}^{2}, m_{h}^{2}, s]$ and  $K[m_{loop}^{2}, m_{h}^{2}, s]$ which have 
the respective limiting values $1/2$, $1/24$ and $1/6$ when $m_{loop}^{2}> > m_{h}^{2}, s$. Hence $A^{i^{\pm}}_{\gamma,Z}$
remain non-vanishing even for infinitely heavy loop masses $m_{loop}$. The $W$ boson contribution, for example, is sensitive to
$I$ and generally dominates over other contributions. In this sense one expects the SM contribution to be large 
compared to the SUSY contributions. This conclusion holds also when both gauge bosons are on their mass shell  \cite{gamma,hunter}.

Therefore, it is convenient to search for an appropriate region of the supersymmetric parameter space where $h\rightarrow
\gamma \gamma$, $h\rightarrow gg$, and $h\rightarrow \gamma \bar{f}f$ can be enhanced due to supersymmetric contributions
\cite{gamma,hunter}. The first two process have been dicussed in \cite{djou} as a function of
relatively large Higgs--sfermion couplings. Here we are mainly concerned with $h\rightarrow \gamma \bar{f}f$ in the strongly
interacting phase of the MSSM where $\alpha_{\tilde{t}}$ is large. The $W$ boson and fermion contributions are common to both
SM and MSSM. We include only sfermions into the discussion as the others (charginos and  charged Higgs) give small
contributions in the parameter space employed here. Among the sfermions the most important contributions follow from scalar 
top quarks as they can be relatively light due to large top Yukawa coupling. Thus, to a good approximation, we represent 
the supersymmetric contributions by $W$, fermion and light stop loops the latter being the pure supersymmetric contribution 
compared to the SM. 

The contribution of the light stop loop has the form
\begin{eqnarray}
A^{\tilde{t}^{\pm}}_{\gamma ,Z}\propto \frac{2 M_{W}}{m_{\tilde{t}_{1}}} \sqrt{\frac{\alpha_{\tilde{t}}}{\alpha_{W}}}
J[m_{\tilde{t}_{1}}^{2}, m_{h}^{2}, s]
\end{eqnarray}
where $\alpha_{W}=g^2/(4 \pi)$. To see the consistency of the light stop dominance, one recalls that large stop left--right mixings
make $\alpha_{\tilde{t}}$ large and $tilde{t}_{1}$ light simultaneously. This causes $A^{\tilde{t}^{\pm}}_{\gamma ,Z}$ to dominate
over other loop contributions, and enhance $R$ significantly. In this sense sfermion contributions, here light stop, can cause
spectacular enhancement in $R$ whereby making supersymmetric contributions observable. 
 
Fig. 1 shows the variation of $R$ with $\alpha_{\tilde{t}}$ for a light $\tilde{t}_{1}$; $m_{\tilde{t}_{1}}=M_{Z}$. Here solid and
dashed curves correspond to $m_{h}=M_{Z}$ and $m_{h}=2 M_{Z}$, respectively. For the given value of $m_{\tilde{t}_{1}}$,
$\alpha_{\tilde{t}}=0.2$ ($\alpha_{\tilde{t}}=1.7$) corresponds to $g_{h\tilde{t}}\sim 300\, \mbox{GeV}$ ($g_{h\tilde{t}}\sim 900\,
\mbox{GeV}$). In this and the next figure we assume a relatively large $\tan\beta$, that is, $\alpha\sim 0$. As is seen from the
figure, $R$ increases monotonically with $\alpha_{\tilde{t}}$ due to the fact that the stop contribution to $R$ increases, as
suggested by the formula (5). However, when the Higgs mass is doubled increase of $R$
is automatically slowed down (dashed curve). Thus, it is more likely to observe the contribution of the stops for a light enough
stop and Higgs. Besides these, one notices that, in general, $R$ is well above unity so that in both cases, despite the
uncertainities in various parameters, it may be quite easy to observe the excess in $R$. Since the stop contribution is able to
dominate over the W--boson contribution, it is quite large compared to the fermion contributions, so that one expects $h g g $
vertex be dominated by stops as well. The upper limit on $\alpha_{\tilde{t}}$ is chosen to be $\sim 1.75$ which is the threshold
value for developing the color--singlet stop bound states via Higgs mediation \cite{canad}. Thus, stop contribution is able to
dominate over those of W and fermions before the onset of the bound state formation. Once the light stop bound state is formed the 
subsequent evolution of the system (depending on it lifetime) could be quite different. Introduction of such bound states 
to the particle spectrum can even replace the properties of the Higgs particles \cite{sasha} falsifying the Higgs search strategy at
the colliders, in particular, at the LHC. 

Figure 2 shows the same quantities in Fig. 1 for a heavier stop $m_{\tilde{t}_{1}}=2\, M_{Z}$. It is seen that both curves
are rescaled according to Fig. 1, that is, now the enhancement in the quantities is much smaller. The given range of
$\alpha_{\tilde{t}}$ again corresponds to the same range for $g_{h\tilde{t}}$ mentioned in the previous paragraph. Since,
$\alpha_{\tilde{t}}$ is below $\sim 0.8$ there is no chance for bound state formation. 

Given the uncertainities in the Higgs masses and huge $\gamma\gamma$ background it is an issue of precision for LHC detectors
to observe the Higgs signal. For this and similar technical reasons one has to consider new mechanisms allowing an easier way of 
detection. In this sense, the final state discussed here constitutes a possible way of obtaining an enhanced single photon signal. 
Observation of the charged particles is not a problem for the experiment, and with the given direction and invariant mass of the 
$\bar{f}f$ pair it might be easier to detect the single prompt photon \cite{prompt}.   

For the process under concern, the interesting thing occurs for larger values of $\alpha_{\tilde{t}}$ for which the threshold 
value for developing stop bound states is exceeded. Once the stops develop bound states they may easily interfere with the Higg
boson signal sought. As was already discussed in \cite{sasha} the stop bound states may develop non-zero vacuum expectation values 
whereby behaving as some component to the Higgs boson signal. In such cases even the existing bounds on the Higgs boson does not
hold and new phenomenological issues arise. In this analysis we have avoided entering this realm of the couplings; however, despite
this MSSM signal may dominate over the SM signal in a wide rage of the fine structur constant $\alpha_{\tilde{t}}$. 

Author thanks to G.Belanger, F.Boudjema, and K.Sridhar for their helpful remarks.

\begin{figure}[htb]
\centerline{
\psfig{figure=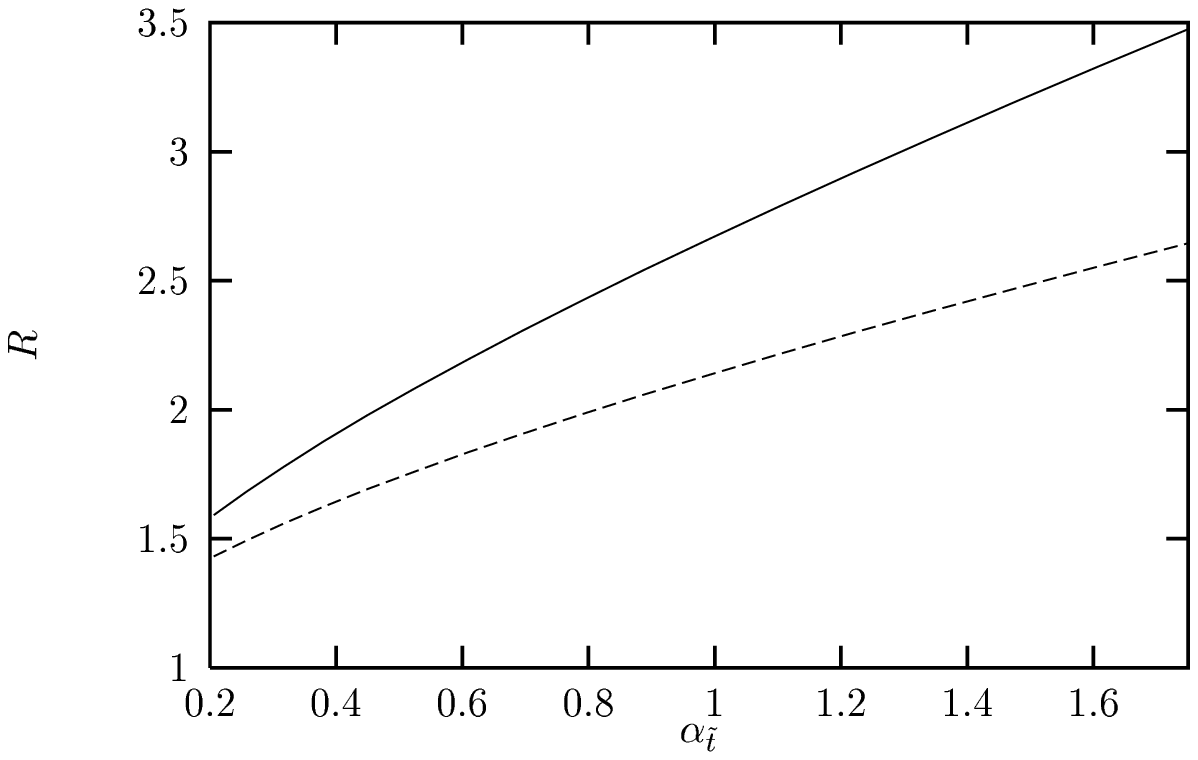,height=10cm,width=9cm,bbllx=4cm,bblly=10cm,bburx=18.cm,bbury=21.cm}}
\caption{\footnotesize Variation of the ratio $R$ in (2) with $\alpha_{\tilde{t}}$ for $m_{\tilde{t}_{1}}=M_{Z}$,
$m_{h}=M_{Z}$ (solid curve) and $m_{h}=2\, M_{Z}$ (dashed curve). $\bar{b}b$ final states are assumed.} 
\end{figure}
\begin{figure}[htb]
\centerline{
\psfig{figure=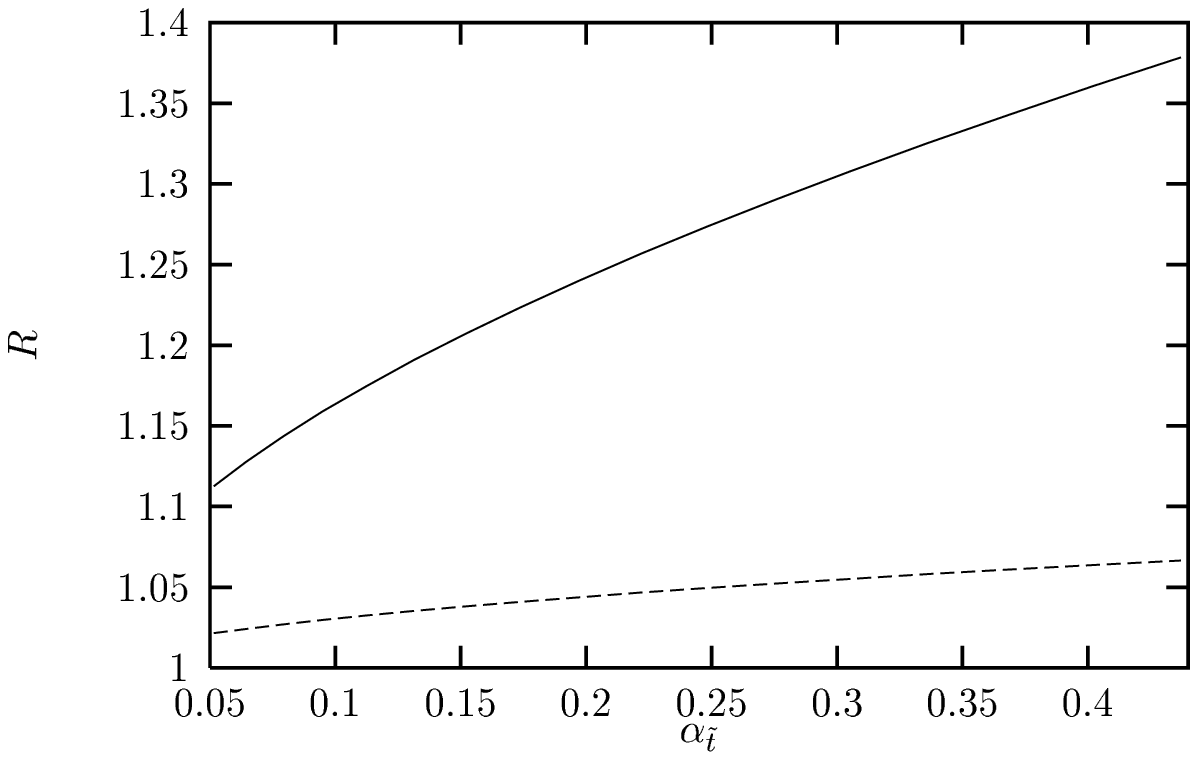,height=10cm,width=9cm,bbllx=4cm,bblly=10cm,bburx=18.cm,bbury=21.cm}}
\caption{\footnotesize Variation of the ratio $R$ in (2) with $\alpha_{\tilde{t}}$ for $m_{\tilde{t}_{1}}=2\, M_{Z}$,
$m_{h}=M_{Z}$ (solid curve) and $m_{h}=2\, M_{Z}$ (dashed curve). $\bar{b}b$ final states are assumed.}
\end{figure}

\begin{thebibliography} {99}
\bibitem{exp} {Particle Data Group Homepage: http://pdg.lbl.gov/.}    
\bibitem{decoupl} {H. E. Haber, CERN-TH/95-109, hep-ph/9505240.} 
\bibitem{upper} {M. Carena, M. Quiros, C. E. M. Wagner, Nucl. Phys. {\bf 
B461} (1996) 407; H. E. Haber, R. Hempfling, A. Hoang, Z. Phys. {\bf C75} 
(1997) 539; J. R. Espinosa, M. Quiros, Phys. Rev. Lett. {\bf 81} (1998) 516.}
\bibitem{lhc} {J. F. Gunion, A. Stange, S. Willenbrock, hep-ph/9602238; 
N. V. Krasnikov, V. A. Matveev, Phys. Part. Nucl. {\bf 28} (1997) 441.}
\bibitem{fusion} {H. Georgi, S. Glashow, M. Machacek, D. Nanopoulos, Phys.
Rev. Lett. {\bf 40} (1978) 692.}
\bibitem{gamma} {T. J. Weiler, T.-C. Yuan, Nucl. Phys. {\bf B318} (1989)
337; T. G. Rizzo, Phys. Rev. {\bf D22} (1980) 178.}
\bibitem{at-cms} {CMS Collaboration, Report CERN-LHCC
94-38; ATLAS Collaboration, Report CERN-LHCC 94-43.}
\bibitem{dreiner} {M. Dittmar, H. Dreiner, Phys. Rev. {\bf D55} (1997) 167.}
\bibitem{others} {D. P. Roy, hep-ph/9803421; C. Kao, hep-ph/9802343; J. 
A. Bagger, hep-ph/9709335; Z. Kunszt, hep-ph/9704263.}
\bibitem{diHiggs} {S. Dawson, S. Dittmaier, M. Spira, hep-ph/9805244; 
T. Plehm, M. Spira, P. M. Zerwas, Nucl. Phys. {\bf 479} (1996) 46; 
M. Spira, A. Djouadi, D. Graudenz, P. M. Zerwas, Nucl. Phys. {\bf B453}
(1995) 17; E. W. N. Glover, J. J. van der Bij, Nucl. Phys. {\bf B309} 
(1988) 282.}
\bibitem{prompt} { P. Aurenche, R. Baier, M. Fontannaz, D. Schiff, Nucl. 
Phys. {\bf B297} (1988) 661.}
\bibitem{tau} {R. K. Ellis, I. Hinchliffe, M. Soldate, J. J. van der 
Bij,  Nucl. Phys. {\bf B297} (1988) 297.}
\bibitem{sasha}{A. Kusenko, V. Kuzmin, I. I. Tkachev, Phys.Lett. {\bf B432}
(1998) 361; G. F. Giudice, A. Kusenko, Phys. Lett. {\bf B439} (1998) 55.}
\bibitem{djou} {A. Djouadi, Phys. Lett. {\bf B435} (1998) 101.}
\bibitem{hunter} {J. F. Gunion, H. E. Haber, G. Kane, S. Dawson,
$The$ $Higgs$ $Hunter's$ $Guide$, Addison-Wesley, New York, (1990).}
\bibitem{canad}{L. D. Leo, J. W. Darewich, Can. J. Phys. {\bf 70} (1992) 412.}
\bibitem{list} {A. Djouadi, J.-L. Kneur, G. Moultaka, hep-ph/9903218;
Phys. Rev.Lett. {\bf 80} (1998) 1830.}
\end{thebibliography}
\end{document}